\documentclass[journal,twoside]{IEEEtran}
\usepackage{cite}
\usepackage{amsmath,amssymb,amsfonts, mathtools}
\usepackage{graphicx}
\usepackage{textcomp}
\usepackage{xcolor}
\usepackage{comment}
\usepackage{commath}
\usepackage{algorithm,algorithmic}
\usepackage{enumerate}
\usepackage{psfrag,subfigure}
\usepackage{amsbsy,epsfig,bbm,mathrsfs,multirow,amsthm}
\usepackage{float}
\usepackage{mathrsfs}
\usepackage{color}
\usepackage{tabulary}
\usepackage{multirow}
\usepackage{cases}
\usepackage{stfloats}
\usepackage{url}
\usepackage[subfigure]{tocloft}
\usepackage{import}
\usepackage{xcolor}

\DeclareMathOperator{\diag}{\mathrm{diag}}
\DeclareMathOperator*{\argmax}{\mathrm{argmax}}
\DeclareMathOperator*{\clip}{\mathrm{clip}}

\begin{document}

\title{RIS-assisted UAV Communications for IoT with Wireless Power Transfer Using Deep Reinforcement Learning}

\author{Khoi Khac Nguyen, Antonino Masaracchia, Tan Do-Duy, H. Vincent Poor, Trung Q. Duong
\thanks{K. K. Nguyen, A. Masaracchia, and T. Q. Duong are with  Queen's University Belfast, Belfast BT7 1NN, UK (e-mail: \{knguyen02, a.masaracchia, trung.q.duong\}@qub.ac.uk).}
\thanks{T. Do-Duy is with HoChiMinh City University of Technology and Education, Vietnam (e-mail: tandd@hcmute.edu.vn)}
\thanks{H. V. Poor is with Princeton University, Princeton, NJ 08544, USA (e-mail: poor@princeton.edu)}
}

\maketitle

\begin{abstract}
Many of the devices used in Internet-of-Things (IoT) applications are energy-limited, and thus supplying energy while maintaining seamless connectivity for IoT devices is of considerable importance. In this context, we propose a simultaneous wireless power transfer and information transmission scheme for IoT devices with support from reconfigurable intelligent surface (RIS)-aided unmanned aerial vehicle (UAV) communications. In particular, in a first phase, IoT devices harvest energy from the UAV through wireless power transfer; and then in a second phase, the UAV collects data from the IoT devices through information transmission. To characterise the agility of the UAV, we consider two scenarios: a hovering UAV and a mobile UAV. Aiming at maximizing the total network sum-rate, we jointly optimize the trajectory of the UAV, the energy harvesting scheduling of IoT devices, and the phaseshift matrix of the RIS. We formulate a Markov decision process and propose two deep reinforcement learning algorithms to solve the optimization problem of maximizing the total network sum-rate. Numerical results illustrate the effectiveness of the UAV's flying path optimization and the network's throughput of our proposed techniques compared with other benchmark schemes. Given the strict requirements of the RIS and UAV, the significant improvement in processing time and throughput performance demonstrates that our proposed scheme is well applicable for practical IoT applications.
\end{abstract}



\section{Introduction}\label{Sec:Intro}
Unmanned aerial vehicles (UAVs) have recently drawn considerable attention due to their agile mobility and cost-effectiveness. UAVs have been used for geometry monitoring, disaster relief \cite{Long:EAI}, emergency services, and wireless networks \cite{Khoi:20:Access}. In wireless networks, UAVs can be deployed at sporting events or in rescue missions to provide and enhance connectivity to the users. UAVs are also used as data collectors that fly to the remote area to collect sensor data \cite{KK:21:TCOM}. However, restrictions regarding flying time and on-board processing ability are bottlenecks that must be dealt with in unexpected environment and complicated missions.

Reconfigurable intelligent surface (RISs) have emerged as a promising technology for future wireless networks. The arrival signal at a RIS is reflected toward the receiver by the RIS's passive elements operated by a module controller. The received signal at the users is composed of elements from the direct channel and the reflective link. It helps to increase the signal quality and reduce the interference. The RIS is usually deployed in high locations such as buildings to reduce the cost of establishing a new station. However, the optimization of RIS performance is still challenging due to the large number of elements and the processing ability of the controller.

One area in which UAVs can be useful is in supporting Internet-of-Things (IoT) applications. Not only can they provide communication coverages, but, since many IoT devices are energy-limited, they can also be sources of power for such devices through downlink power transfer. A downlink power transfer and uplink information transmission protocol can implemented in two phases: wireless power transfer (WPT) and wireless information transmission (WIT). In the first phase, the IoT devices harvest energy from a base station (BS) or from the UAV. The harvested energy is then used for transmitting local information to receivers or back to the UAV and the BS. By using such a downlink power transfer and uplink information transmission protocol, the IoT devices can obtain the energy to establish and maintain communication with the BS and the UAV. 

Machine learning is an effective tool for optimizing the performance of large-scale networks under dynamic environments. One of the approaches is deep reinforcement learning (DRL), which is a combination of reinforcement learning and neural networks. In wireless networks, DRL algorithms are used for maximizing the network performance, reducing power consumption and improving the processing time for real-time applications \cite{KK:19:Access, Khoi:19:Access, Khoi:20:Access}. DRL algorithms are powerful in wireless networks because the agents do not need pre-collected data for training. Rather, DRL agents interact with their environment and establish training samples for the responses in those interactions. The neural networks are trained by up-to-date state transitions to adjust their parameters for maximizing a designated reward. Then, the trained networks are deployed for real-time prediction.

\subsection{State-of-the-art}
UAV-assisted wireless communications have been widely used to enhance network coverage as well as network performance \cite{Long:EAI, LN:19:SPAWC, Khoi:20:Access}. In \cite{Long:EAI, LN:19:SPAWC}, the authors used the UAV for providing the network for the disaster relief missions. A UAV can also serve as an energy source provider for device-to-device communications \cite{Khoi:20:Access}.
Recently, RIS technology has been introduced as a low-cost and easily installed technology to mitigate interference and direct transmitted signals toward their receivers \cite{EB:19:Access, SA:20:TC, HY:20:JSAC, YL:21:TC}. In \cite{SA:20:TC}, the authors considered two-way communications assisted by a RIS. The reciprocal channel to maximize the signal-to-interference-plus-noise ratios (SINR) and the non-reciprocal channel with the target of maximization of the minimum SINR was considered. The gamma approximation was used for the reciprocal channel, while the semidefinite programming relaxation and a greedy-iterative method were used for the non-reciprocal channel. In \cite{HY:20:JSAC}, an iterative algorithm with low computation complexity was proposed to solve the joint optimization of transmit beamforming vector and the phase shift of a RIS under proper and improper Gaussian signalling. In \cite{YL:21:TC}, the authors optimized the beamforming matrices at the BS and the reflective vector at the RIS to minimize the total transmit power at a multiple-input-single-output (MISO) non-orthogonal multiple access (NOMA) networks. An algorithm based on the second-order cone programming-alternating direction method of multipliers was proposed to reach an optimal local problem.

By utilising both advantages of the UAV and the RIS, the received signal at the ground users is strengthened while the power consumption is reduced and the flying time of the UAV can be extended \cite{LG:20:Access, SL:20:WCL, AR:21:IOT, KK:21:NCE}. In \cite{SL:20:WCL}, the UAV's trajectory and the RIS's passive beamforming vector were optimized to maximize the average rate in RIS-assisted UAV communications. The problem was derived into two subproblems; then, a closed-form phase shift algorithm was introduced to find the local optimal reflective matrix and the successive convex approximation was used to find the suboptimal trajectory solution. In \cite{AR:21:IOT}, the UAV acts as a mobile relay and the RIS was used to provide short packets communications ultra-reliable and low-latency between ground transmitter and ground IoT devices. The UAV's position, the RIS phase shift and the blocklength were optimized to minimize the total decoding error rate by using a polytope-based method, namely Nelder-Mead simplex.

Along with the development of the IoT devices is the increase power supply for each device. However, not all the nodes are equipped with fixed power providers and have solar batteries. Thus, we need to find a solution to provide power to the nodes. The downlink power transfer and uplink information transmission protocol is one of the solutions to enable the IoT devices to harvest energy from source providers and switch to information transmission in the uplink phase on demand \cite{YZ:20:VT, CH:20:JSAC, Pan:20:JSAC, HY:21:SP, SL:21:WC}. That helps reduce the power consumption as well as the cables, wires for providing power. In \cite{YZ:20:VT}, the authors designed a time-switching protocol for a RIS with the energy harvesting phase to charge the RIS capacitor and the signal reflecting phase to assist the transmission from the access point (AP) to the receivers. The AP's transmit beamforming, the RIS's phase scheduling and the passive beamforming were optimized to maximize the information rate. The resultant two sub-problems were solved following the conventional semidefinite relaxation method and monotonic optimization. In \cite{CH:20:JSAC}, the transmit precoding matrices of the BS and the RIS's passive phase shift matrix were optimized for maximizing the weight sum-rate of all information receivers in a power transfer scenarios.

The demand for a technique that is flexible and adaptive to changes of the environment while satisfying real-life constraints is rising, and DRL algorithms are among the most potential methods to deal with these problems in wireless networks \cite{KK:19:Access, Khoi:19:Access, Khoi:20:Access, KK:21:TCOM}. Recently, DRL algorithms are also used for the RIS-assisted wireless networks and have shown promising results \cite{YC:20:WC, LW:20:arvix, HY:21:WC, KF:20:WCL, KK:21:NCE}. The power allocation and the phase shift optimization were optimized for maximizing the sum rate in \cite{YC:20:WC}. In \cite{LW:20:arvix}, a RIS-assisted UAV was deployed for serving ground users. The trajectory and phase shift optimization relying on DRL for maximizing the sum rate and fairness of all users was proposed. In \cite{HY:21:WC}, the authors used a RIS to assist the secure communications against eavesdroppers. The DRL algorithms were used to optimize the BS beamforming and the RIS's reflecting beamforming were shown to improve the secrecy rate and the quality-of-service satisfaction probability. In \cite{KF:20:WCL}, deep deterministic policy gradient was proposed to obtain the optimal phase shift matrix at the RIS to maximize the received signal-to-noise ratio (SNR) in a MISO system. In \cite{KK:21:NCE}, the joint optimization of the power and the RIS's phase shift in a multi-UAV-assisted network is considered.


\subsection{Contributions}
Inspired by the aforementioned discussion, in this paper, we consider the IoT wireless networks with the support of an UAV, and one RIS, and employ the downlink power transfer and uplink information transmission protocol for maximizing the total network's sum-rate. In particular, we adopt the harvest-then-transmit protocol, which means the IoT devices use all the harvested energy in the first phase for transmitting during the remaining time. Then, two DRL algorithms are deployed for solving the problem in RIS-assisted UAV communications. In summary, our main contributions are as follows:
\begin{itemize}
	\item We conceive a system model of UAV-assisted IoT wireless power transfer with the support of a RIS. The IoT devices harvest energy in the downlink and transmit information in the uplink to the UAVs.
	\item To characterise the agility of UAV in supporting the energy harvesting (EH) and information transmission of IoT devices, we consider two scenarios of UAV. Firstly, the UAV is hovering at the centre of the cluster and provides energy to the IoT devices. The RIS helps alleviate the uplink interference when the IoT devices transmit their information to the UAV. Secondly, the UAV is deployed in an initial location and required to find a better location for communication. In each location of the UAV's flying trajectory, the EH time scheduling and the RIS's phase shift matrix are optimized for maximizing the network throughput performance.
	\item For the defined problem, we formulate a Markov decision process (MDP) \cite{BD:95:Book:v1} with the definition of the state space, action space and the reward function. Then, we propose a method based on deep deterministic policy gradient (DDPG) and proximal policy optimization algorithm (PPO) for solving the maximization game.
	\item Our results suggest that with the support of the RIS, a better connection is established and the overall performance is significantly improved.
\end{itemize}


\section{System Model and Problem Formulation}\label{Sec:Model}
We consider that the system includes one single-antenna UAV and $N$ ground IoT devices distributed randomly. However, there are some practical scenarios where IoT devices are located in a crowded area with surrounding obstacles and objects. In such complex environment, IoT devices suffer high attenuation and severe path loss. In this case, the RIS is also installed at the wall of a tall building to enhance the communication quality by reflecting signal from the UAV to the IoT devices. Here, we deploy a RIS composed of $K$ elements to enhance the network performance. The 3D coordinate of the UAV at the time step $t$ is $X^t_{UAV} = (x^t_{UAV}, y^t_{UAV}, z^t_{UAV})$. In this paper, we consider the fixed attitude of the UAV at $H_{UAV}$. The location of the $n$th IoT devices at time step $t$ is $X^{t}_n = (x^{t}_n, y^{t}_n)$ with $n = 1, \dots , N$. The position of the RIS component $k \in K$ at time step $t$ is $(x^t_k, y^t_k, z^t_k)$. In this paper, we use the wireless downlink power transfer and uplink information transmission protocol for deploying the UAV and collecting data. Particularly, we have two phases: wireless power transfer (WPT) and wireless information transmission (WIT). In the first phase, the downlink is activated to transfer energy to the IoT devices from the UAV during time span $\tau\mathcal{T}$. Then, the WIT phase takes place when the IoT devices transmit information to the UAV in the uplink during $(1-\tau) \mathcal{T}$. We normalise the length of time step to $\mathcal{T} = 1$ for convenience.

\begin{figure}[h!]
	\centering
	\subfigure{\includegraphics[width=0.5\textwidth]{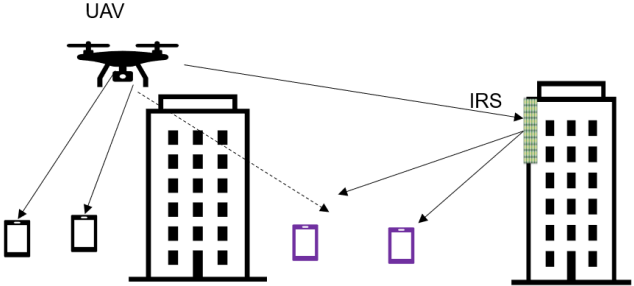}}
	\caption{System model of UAV-assisted IoT wireless communications with the support of a RIS.}
	\label{fig:System}
\end{figure}

\subsection{Channel model}
We denote the channel gain between the UAV and the RIS, between the RIS and the $n$th IoT device, and the direct link from the UAV to $n$th IoT node at time step $t$ by $H^t \in \mathbb{C}^{1\times K}, h_{RIS, n}^t \in \mathbb{C}^{1 \times K}$, and $h_n^t$, respectively. The small-scale fading of the direct link from the UAV to the IoT devices is assumed to be Rayleigh fading due to the extensive scatters. The air-to-air channel is considered for the UAV and the RIS link, while the link from the RIS to the IoT devices can be modelled by the Rician fading channel. 

The distance between UAV and the $k$th RIS in time step $t$ is given by
\begin{equation}
d_k^t = \sqrt {(x_{UAV}^t -x_k^t)^2 + (y_{UAV}^t - y_{UAV}^t)^2 + (z^t_{UAV} - z^t_k)^2}.
\end{equation}
Similarly, we denote the distance between the UAV and the $n$th IoT device and between the $k$th RIS element and the $n$th IoT node by $d_n^t$ and $d_{k,n}^t$, respectively. 

The channel gain between the UAV and the $n$th IoT device is given by
\begin{equation}
h_n^t = \sqrt{\beta_0 (d_n^t) ^ {-\kappa_1} } \hat{h},
\end{equation}
where $\beta$ and $\kappa_1$ are the path loss at reference distance $1m$ and the path loss exponent for the UAV and the IoT devices link, respectively;  $\hat{h}$  represents the small-scale fading modelled by complex Gaussian distribution with zero-mean and unit-variance $\mathcal{C} \mathcal{N} (0, 1)$.

Similarly, the channel gain between the UAV and the RIS is an air-to-air channel dominated by the line-of-sight (LoS) links. Thus, the channel of the UAV-RIS link in time step $t$ is denoted as follows:
\begin{equation}
\begin{split}
H^t = \sqrt{\beta_0(d_k^t)^{- \kappa_2}} \big[ 1,& e^{-j \frac{2\pi}{\lambda}d \cos(\phi^t_{AoA})}, \dots,\\ & e^{-j \frac{2\pi}{\lambda}(K -1) d \cos(\phi^t_{AoA})} \big]^T,
\end{split}
\end{equation}
where the right term is the array signal from the UAV to the RIS, $\cos(\phi^t_{AoA})$ is the cosine of the angle of arrival (AoA) from the UAV to RIS; $\kappa_2$, $d$ and $\lambda$ are the path loss exponent for the UAV and the RIS link, the antenna separation and the carrier wavelength, respectively.

The channel gain between the RIS and the $n$th IoT device following the Rician fading is expressed as
\begin{equation}
h_{RIS,n}^t = \sqrt{\beta_0 (d_{k,n}^t)^{-\kappa_3} } \big( \sqrt{ \frac{\beta_1}{1+\beta_1}} h^{LoS}_{RIS,n} + \sqrt{\frac{1}{\beta+1}} h^{NLoS}_{RIS,n} \big),
\end{equation}
where the deterministic LoS component is denoted by $ h^{LoS}_{RIS,n} =
[ 1, e^{-j \frac{2\pi}{\lambda}d\cos(\phi^t_{AoD})}, \dots, e^{-j\frac{2\pi}{\lambda}(K-1)d \cos(\phi^t_{AoD})} ]$ and the non-line-of-sight (NLoS) component is the Rayleigh fading that follows the complex Gaussian distribution with zero mean and unit variance;  $\cos{\phi_{AoD}}$ is the angle of departure (AoD) from the RIS to IoT devices; $\beta_1$ is the Rician factor, and $\kappa_3$ is the path loss exponent for the RIS and IoT devices link.

\subsection{Power transfer phase}
The achievable signal at the $n$th IoT device is composed of direct signal from the UAV and the reflected signal from the RIS at time step $t$ as
\begin{equation}
y_{1n}^t = (h_n^t + H^t \Phi^t h_{RIS, n}^t)\sqrt{P_0}x + \varrho^2,
\end{equation}
where $\varrho^2$ is the noise signal following the complex Gaussian distribution $\mathcal{C} \mathcal{N} (0, \alpha^2)$, $x$ is the symbol signal from the UAV and $P_0$ is the transmission power at the UAV; $\Phi^t = \diag[\phi_1^t, \phi_2^t, \dots, \phi^t_K]$ is the diagonal matrix at the RIS, where $\phi_k^t = e^{j\theta^t_k}, \forall k = 1, 2, \dots, K$ and $\theta_k^t \in [0, 2\pi]$ denotes the phase shift of the $k$th element in the RIS at time step $t$.

In the WPT phase, the UAV transfers energy to the IoT devices during time span $\tau^t$ at time step $t$. Thus, the received power at the $n$th IoT devices at time step $t$ is given by
\begin{equation}
p_n^t = \tau^t \eta P_0 |h_n^t + H^t \Phi^t g_n^t|^2,
\end{equation}
where $\eta$ is the power transfer efficiency.

\subsection{Information transmission phase}
We assume that the IoT devices do not have fixed energy sources and use all the harvested energy for the WIT phase. The signal received at the UAV from the $n$th IoT devices is given by
\begin{equation}
y_{2n}^t = (h_n^t + H^t \Phi^t h_{RIS, n}^t)\sqrt{p_n}u_n + \varrho^2,
\end{equation}
where $u_n$ is the symbol signal from the $n$th IoT devices to the UAV. The received SINR at the UAV from transmission of the $n$th IoT device  at time step $t$ can be formulated as follows:
\begin{equation}
\gamma_n^t = \frac{p_n^t |h_n^t + H^t \Phi^t g_n^t|^2}{\sum_{m \ne n}^N p_m^t |h_m^t + H^t \Phi^t g_m^t|^2 + \alpha^2},
\end{equation}

The sum-rate from the IoT devices at time step $t$ is formulated as follows:
\begin{equation}
R_{total}^t = \sum_{n=1}^N (1-\tau^t) B \log_2 (1 + \gamma_n^t),
\end{equation}
where $B$ is the bandwidth.

Our objective is to maximize the achieved sum-rate performance by optimizing the phase shift matrix $\Phi$ at the RIS, the UAV's trajectory $\varGamma$ and the EH time $\tau$ as
\begin{equation}\label{equ:EE}
\begin{split}
\max_{\tau, \Phi, \varGamma} & \sum_{n=1}^N (1-\tau^t) B \log_2 (1 + \gamma_n^t)\\
s.t. \quad & 0 < \tau < 1 \\
& \theta^k \in [0, 2 \pi], \forall k \in K \\
& v \le v_{max}\\
& X_{UAV} \in Z\\
\end{split}
\end{equation}
where $Z$ represents the flying restricted area in the vertical and horizontal dimensions; $v$ and $v_{max}$ are the velocity and the maximum flying velocity of the UAV, respectively.

\section{Hovering UAV for downlink power transfer and uplink information transmission in RIS-assisted UAV communications}\label{Sec:StaticUAV}
Besides WPT, the UAV uses most energy for its movement. Thus, to extend the operating time, the UAV is considered to hover at a fixed position at the central of the cluster. For this scenario, we formulate the MDP \cite{BD:95:Book:v1} by a 4-tuple $< \mathcal{S}, \mathcal{A}, \mathcal{P}, \mathcal{R} >$ where $\mathcal{S}, \mathcal{A}$ are is the agent's state space and action space, respectively; $\mathcal{P}_{ss'}(a)$ is the state transition probability with $s = s^t, s' =s^{t+1} \in \mathcal{S}$, $a \in \mathcal{A}$; $\mathcal{R}$ is the reward function. Then, we formulate a game to solve the problem in Equ. (\ref{equ:EE}).
\begin{itemize}
	\item{\emph{Agent}}: The centralised processor will act as an agent. The agent interacts with the environment to find an optimal policy $\pi^*$ for maximizing the total sum-rate. After training, the action-making schemes will be deployed to the UAV to predict the proper EH time scheduling $\tau$ and the RIS can choose the phase shift matrix $\Phi$.
	
	\item{\emph{State space}}: The channel is composed of both direct link and the reflective channel. Thus, we define the state space as
	\begin{equation}\label{MDP:state}
	\mathcal{S} = \{ h_1 + H \Phi g_1, h_2 +H \Phi g_2, \dots, h_N + H \Phi g_N \} ,
	\end{equation}
	In time step $t$, the UAV has the state $s^t = \{h_1^t +H^t \Phi^t g_1^t, h_2^t + H^t \Phi^t g_2^t , \dots,  h_N^t +H^t \Phi^t g_N^t \}$
	
	\item{\emph{Action space}}: The UAV hovers at a fixed position; thus, we optimize the EH time $\tau$ and the RIS's phase shift $\Phi$. The action space is defined as
	\begin{equation}
	\mathcal{A} = \{\tau, \theta_1, \theta_2, \dots, \theta_K \}
	\end{equation}
	At the state $s^t$, the UAV takes the action $a^t = \{  \tau^t, \theta_1^t, \theta_2^t, \dots, \theta_K^t \}$ and move to the next state $s' = s^{t+1}$.
	\item{\emph{Reward function}}: The UAV interacts with the environment to find the maximum obtained reward. In our work, we formulate the reward function to obtain the maximum total sum-rate performance as
	\begin{equation}\label{MDP:reward}
	\mathcal{R} =\sum_{n=1}^N (1-\tau^t) B \log_2 (1 + \gamma_n^t)
	\end{equation}
\end{itemize}

The UAV is hovering at $X_{UAV}$ and chooses the action $a^t$ based on the achieved channel state information (CSI). Then, the UAV transfers the energy during $\tau$ to the IoT devices and the RIS controller adjusts the phase shift for each element. During the remaining time $(1-\tau)$, the RIS will not change the phase shift while the IoT devices transmit information in the uplink to the UAV. It is challenging while the RIS plays a crucial role in mitigating the interferences. Thus, we need to find an intelligent scheme for the RIS to maximize the network performance in the downlink power transfer and uplink information transmission protocol. We propose a DRL, namely DDPG algorithm, to find an optimal policy for the UAV and the RIS.

The DDPG algorithm is a hybrid model composed of the value function and policy search methods. Thus, the DDPG algorithm is suitable for large-scale action and state spaces. Based on the current policy, the actor function $\mu(s; \theta_\mu)$ maps the states to a specific action with $\theta_\mu$ being the actor network parameters, while the critic function $Q(s, a)$ evaluates the quality of the action taken. In the DDPG algorithm, we use \emph{experience replay buffer} and \emph{target network} technique to improve the convergence speed and avoid excessive calculations.

The agent iteratively interacts with the environment by executing the action $a^t$ and receives the response with instant reward $r^t$ and the next state $s^{t+1}$. The tuple of $(s^t, a^t, r^t, s^{t+1})$ is then stored in a replay buffer $D$ for training the actor and critic network. The buffer $D$ is updated by adding new samples and discarding the oldest ones due to its finite size setting. After achieving enough samples, the agent takes a batch $B$ of transitions for training the network. Particularly, we train the actor and critic network using stochastic gradient descent (SGD) over a mini-batch $B$ samples.

The state-value function $V$ is defined by following the policy $\pi$ at the state $s$ as follows:
\begin{equation}
V^\pi = \mathbb{E} \Big \{\mathcal{R}|s, \pi \Big\},
\end{equation}
where $\mathbb{E}$ is the expectation operation.

The state-action value $Q$ is obtained when the agent at the state $s$ takes action $a$ following the policy $\pi$ as follows:
\begin{equation}
Q ^\pi (s, a) = \mathbb{E} \Big(r(s, a)\Big) + \zeta \sum_{s' \in \mathcal{S}}{P_{ss'}(a) V(s')}.
\end{equation}

Let us denote the parameters of the critic network and the target critic network by $\theta_{q}$ and $\theta_{q'}$, respectively. The critic network is updated by minimizing
\begin{equation}
\label{equ:loss}
L = \frac{1}{B}\sum_{i}^B \Bigg(y^i - Q(s^i, a^i; \theta_{q}) \Bigg)^2,
\end{equation}
with
\begin{equation}
\begin{split}
y^i = r^i(s^i,a^i) + \zeta Q'(s^{i+1}, a^{i+1}; \theta_{{q'}})|_{a^{i+1} = \mu'(s^{i+1}; \theta_{\mu'})}.
\end{split}
\end{equation}
where the action at time step $(i+1)$ can be obtained by running the target actor network $\mu'$ with the state $s^{i+1}$; $\theta_{\mu'}$ denotes the parameters of the target actor network and $\zeta$ is the discounting factor.

The actor network parameters are updated by
\begin{equation}
\label{equ:updateActor}
\nabla_{\theta_{\mu}}J \approx \frac{1}{B}\sum_{i}^B \nabla_{a^i} Q(s^i, a^i; \theta_{q})|_{a^i=\mu(s^i)} \nabla_{\theta_{\mu}} \mu(s^i; \theta_{\mu}).
\end{equation}

Moreover, we duplicate the actor network and the critic network after a number of episodes to create a target actor and a target critic network. It helps reduce the excessive calculations by using only one network to estimate the target value. The target actor network parameters $\theta_{q}$ and the target critic network parameter $\theta_{\mu'}$ are updated by using soft target updates associated with $\varkappa \ll 1$
\begin{equation}
\label{equ:updatePar1}
\theta_{q'} \leftarrow \varkappa\theta_{q} + (1-\varkappa) \theta_{q'},
\end{equation}
\begin{equation}
\label{equ:updatePar2}
\theta_{\mu'} \leftarrow \varkappa \theta_{\mu} + (1-\varkappa) \theta_{\mu'}.
\end{equation}

For \emph{explorations} and \emph{exploitations} purpose, we add a noise process of $\mathcal{N}(0,1)$ as follows \cite{Lillicrap:15}:
\begin{equation}
\mu'(s^t) = \mu(s^t; \theta^t_{\mu}) + \psi \mathcal{N}(0,1),
\end{equation}
where $\psi$ is a hyper-parameter. In this section, we assume the UAV is hovering at a fixed position to reduce the flying energy consumption. It is a trade-off game with the energy and total achievable sum-rate. In the next section, we propose a joint optimization of trajectory, EH time and the phase shift to maximize the network throughput in a short operation time.

\section{Joint trajectory, EH time scheduling and the RIS phase shift optimization using deep reinforcement learning}\label{Sec:DDPG}
Given a short flying time of the UAV, to maximize total achievable sum-rate, we propose a joint optimization scheme between the UAV's trajectory, EH time scheduling of IoT, and the RIS's phase shift. We define the state space and the reward function as in Section \ref{Sec:StaticUAV}. We modify the action space as follows:
\begin{equation}
\mathcal{A} = \{ v , \varsigma, \tau, \theta_1, \theta_2, \dots, \theta_K \}
\end{equation}

At the state $s^t$, the UAV takes the action $a^t = \{  v^t , \varsigma^t, \tau^t, \theta_1^t, \theta_2^t, \dots, \theta_K^t \}$ and moves to the next state $s' = s^{t+1}$. Particularly, the position of the UAV at time step $(t+1)$ is represented as follows:
\begin{equation}
X^{t+1}_{UAV} = \left \{ \begin{array}{rcl}
x^{t+1}_{UAV} = &x^t_{UAV} +  v^t  \cos{\varsigma^t}+ \Delta x^{t+1}\\
y^{t+1}_{UAV} = &y^t_{UAV} +  v^t \sin{\varsigma^t} + \Delta y^{t+1}\\
H^{t+1}_{UAV} = &H^t_{UAV} + \Delta H^{t+1}, \\
\end{array} \right.
\end{equation}
where $\Delta x^{t+1}, \Delta y^{t+1}, $ and $\Delta H^{t+1}$ are the environmental noise on the UAV at time step $(t+1)$. The UAV is flying from the position $X^{t}_{UAV}$ to $X^{t+1}_{UAV}$ but still needs to satisfy the flying zone constraint $X_{UAV} \in Z$. Moreover, the velocity of the UAV is set to satisfy the requirement $v \le v_{max}$ and the flying angle is set to by a constraint, $\varsigma \in [0, 2\pi]$.

Our objective is to find the optimal policy $\pi^*$ for maximizing the expected reward $\mathcal{R}$. The agent has the local knowledge and interacts with the environment to receive the reward. Base on the received reward, the agent adjusts the policy $\pi$ and executes a new action at a new state. The agent can find a better policy with a better reward by the iterative interactions. After each execution of the action, the UAV will move to a new position and receive responses from the environment. By interacting iteratively with the environment, the agent can choose the proper velocity and the flying direction for the UAV in each time step based on the achieved CSI. Simultaneously, the EH scheduling $\tau$ and the phase shift matrix are also optimized for maximizing network performance. Here, $M$ and $T$ are the number of the maximum episodes and time steps, respectively. The details of our DDPG algorithm-based technique for joint trajectory design, EH time and phase shift matrix optimization in RIS-assisted UAV communications are presented in Alg. \ref{alg:DDPG}.

\begin{algorithm}[t!]
	\caption{Deep deterministic policy gradient algorithm for joint trajectory design, EH time and phase shift optimization in RIS-assisted UAV communications}
	\begin{algorithmic}[1]
		\label{alg:DDPG}
		\STATE Initialise the actor network $\mu(s; \theta_\mu)$, target actor network $\mu'$ and the critic network $Q(s, a; \theta_q)$, the target critic networks $Q'$.
		\STATE Initialise replay memory pool $\mathcal{D}$
		\FOR{episode = $1,\dots, M$}
		\STATE Initialise an action exploration process $\mathcal{N}$
		\STATE Receive initial observation state $s^0$
		\FOR{iteration = $1,\dots, T$}
		\STATE Find the action $a^t$ for the state $s^t$
		\STATE Execute the action $a^t$
		\STATE Update the reward $r^t$ according to (\ref{MDP:reward})
		\STATE Observe the new state $s^{t+1}$
		\STATE Store transition $(s^t, a^t, r^t, s^{t+1})$ into replay buffer $\mathcal{D}$
		\STATE Sample randomly a mini-batch of $B$ transitions $(s^i, a^i, r^i, s^{i+1})$ from $\mathcal{D}$
		\STATE Update critic parameter by SGD using the loss (\ref{equ:loss})
		\STATE Update the actor policy parameter (\ref{equ:updateActor})
		\STATE Update the target networks as in (\ref{equ:updatePar1}) and (\ref{equ:updatePar2})
		\STATE Update the state $s^t_i = s^{t+1}_i$
		\ENDFOR
		\ENDFOR
	\end{algorithmic}
\end{algorithm}

\section{Proximal policy optimization technique for joint trajectory, EH time and the phase shift optimization}\label{Sec:PPO}
For the continuous state and action space as in our problem, we propose an on-policy algorithm, namely the PPO algorithm, for the joint optimization of trajectory, EH time and the phase shift of the RIS. We define the policy by $\pi$ with the parameter $\theta_{\pi}$. Here, we train the policy and adjust the parameter to find an optimal policy $\pi^*$ by running the SGD over a mini-batch of $B$ transitions $(s^i, a^i, r^i, s^{i+1})$. The policy parameters are updated for optimizing the objective function as follows:
\begin{equation}
\theta_{\pi}^{i+1} = \argmax_{\theta_{\pi}} \frac{1}{B}\sum_{i}^B \nabla_{a^i} \mathcal{L}(s^i, a^i; \theta_{\pi}).
\end{equation}

In the PPO algorithm, the agent interacts with the environment to find the optimal policy $\pi^*$ with the parameter $\theta_{\pi^*}$ that maximizes the reward as
\begin{equation}
\begin{split}
\mathcal{L} (s, a; \theta_{\pi}) = \mathbb{E} \Bigg[  p^t_\theta A^\pi(s, a) \Bigg],
\end{split}
\end{equation}
where $p^t_\theta =\frac{\pi(s, a; \theta_{\pi})}{\pi (s, a; \theta_{old})} $ is the probability ratio of the current policy and previous policy; $A^\pi (s, a)$ is the advantage function \cite{JS:16:ICLR}.

Here, if we use only one network for the policy, the excessive modification occurs during the training stage. Thus, we use the clipping surrogate method as follows \cite{JS:17:PPO}:
\begin{equation}
\begin{split}
\mathcal{L}^{\clip} (s, a; \theta_{\pi}) = \mathbb{E} \Bigg[ & \min \Big(p^t_\theta A^\pi(s, a), \\& \clip(p^t_\theta, 1-\epsilon, 1+\epsilon )A^\pi (s, a)\Big) \Bigg],
\end{split}
\end{equation}
where $\epsilon$ is a small constant. In this paper, the advantage function $A^\pi(s, a)$ \cite{Mnih:16} is formulated as follows:
\begin{equation}
\label{equ:A}
A^\pi(s, a) = r^t + \zeta V^\pi(s^{t+1}) -V^\pi(s^t).
\end{equation}

The policy is then trained by a mini-batch $B$ and the parameters are updated by
\begin{equation}
\label{equ:policyPPO}
\theta^{i+1} = \argmax_{\theta_{\pi}} \mathbb{E} \Big[\mathcal{L}^{\clip}(s, a ; \theta_{\pi})\Big].
\end{equation}

The details of our PPO algorithm-based technique for joint trajectory design, EH time and phase shift matrix optimization in RIS-assisted UAV communications are presented in Alg. \ref{alg:PPO}.

\begin{algorithm}[t!]
	\caption{Our proposed approach based on the PPO algorithm for the RIS-assisted UAV communications}
	\begin{algorithmic}[1]
		\label{alg:PPO}
		\STATE Initialise the policy $\pi$ with the parameter $\theta_\pi$
		\STATE Initialise the penalty method parameters $\epsilon$
		\FOR{episode = $1,\dots, M$}
		\STATE Receive initial observation state $s^0$
		\FOR{iteration = $1,\dots, T$}
		\STATE Find the action $a^t$ based on the current state $s^t$ by following the current policy
		\STATE Execute the action $a^t$
		\STATE Update the reward $r^t$ according to (\ref{MDP:reward})
		\STATE Observe the new state $s^{t+1}$
		\STATE Update the state $s^t_i = s^{t+1}_i$
		\STATE Collect set of partial trajectories with $B$ transitions
		\STATE Estimate the advantage function as (\ref{equ:A})
		\ENDFOR
		\STATE Update policy parameters using SGD with a mini-batch $B$ of the collected samples
		\begin{equation}
		\theta^{i+1} = \argmax_{\theta_\pi} \frac{1}{B} \sum^{B} \mathcal{L}^{\clip}(s, a ; \theta_\pi)
		\end{equation}
		\ENDFOR
	\end{algorithmic}
\end{algorithm}

\section{Simulation Results}\label{Sec:Results}
In our works, we use the Tensorflow 1.13.1 \cite{Abadi:16} for implementing our algorithms. We deploy the UAV at $(0, 0, 200)$, the RIS at $(200, 0, 50)$ and assume $d/\lambda = 1/2$ for convenience. All other parameters are provided in Table \ref{tab:Params}. In order to compare our proposed model with other baseline schemes, in this paper, we consider the techniques as follows:
\begin{itemize}
	\item \textbf{optimization with the hovering UAV}: the UAV is maintained at a fixed position at the centre of the cluster $(0, 0, H_{UAV})$. We optimize the EH time $\tau$ and the phase shift matrix at the RIS. We use the DDPG algorithm (H-DDPG) and the PPO algorithm (H-PPO) for the problem in the hovering UAV scenario.
	\item \textbf{Our proposed model with mobile UAV}: For the game formulated as in Section \ref{Sec:DDPG}, we use the DDPG algorithm (F-DDPG) and the PPO algorithm (F-PPO) for solving the problem of joint optimization of trajectory, EH time scheduling and the phase shift matrix at the RIS.
	\item \textbf{Random selection scheme (RSS)}: The value of $\Phi$ is selected randomly and we use the DDPG algorithm (RSS-HDDPG) for optimizing the EH time $\tau$ in the hovering UAV scenario.
	\item \textbf{Random EH time (REH)}: The EH time $\tau$ is selected randomly and the flying path and the phase shift $\Phi$ are optimized to maximize the performance. We use the DDPG algorithm (REH-DDPG) and the PPO algorithm (REH-PPO) for optimization.
	\item \textbf{Without RIS}: We do not deploy the RIS in this scenario and optimize the EH time $\tau$ in the hovering UAV scenario using the DDPG algorithm (WithoutRIS-HDDPG), PPO algorithm (WithoutRIS-HPPO).
\end{itemize}

\begin{table}[h!]
	\renewcommand{\arraystretch}{1.2}
	\caption{SIMULATION PARAMETERS}
	\label{tab:Params}
	\centering
	\begin{tabular}{l|l}
		\hline
		Parameters & Value \\
		\hline
		Bandwidth ($W$)  & $1$ MHz \\
		UAV transmission power & $5$ W \\
		UAV maximum speed per timestep &$20$ m\\
		Path-loss parameter & $\kappa_1 = 4, \kappa_2 = 2, \kappa_3 = 2.2$\\
		Channel power gain & $\beta_0 = -30$ dB\\
		EH efficiency & $\eta = 0.5$ \\
		Rician factor & $\beta_1 = 4$\\
		Noise power & $\alpha^2 = -134$ dBm\\
		Clipping parameter & $\epsilon = 0.2$\\
		Discounting factor & $\zeta = 0.9$\\
		Max number of IoT devices & $20$\\
		Initial batch size & $ K = 32$ \\
		
		\hline
	\end{tabular}
\end{table}

\begin{figure}[t!]
	\centering
	\subfigure{\includegraphics[width=0.5\textwidth]{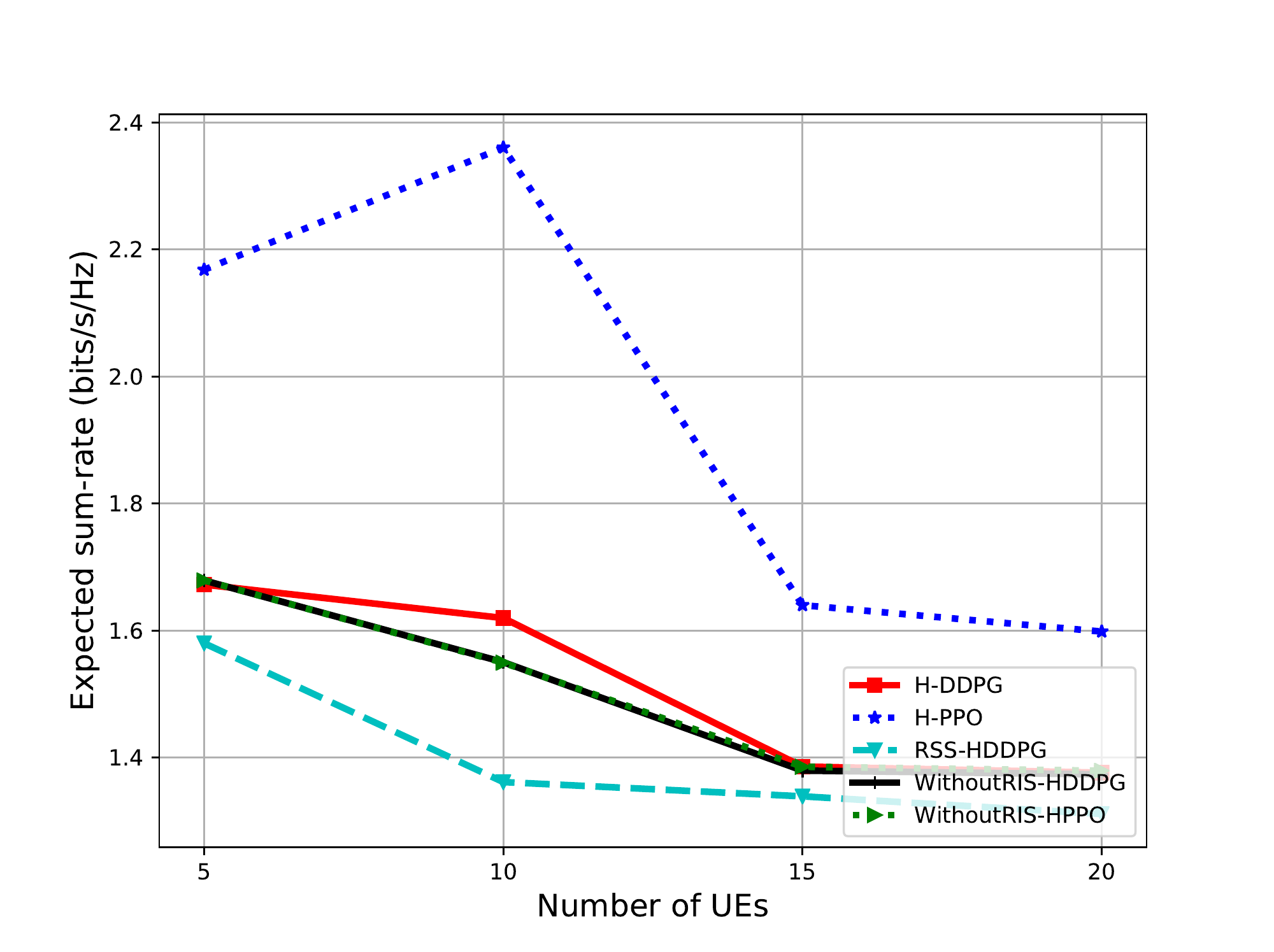}}
	\caption{The sum-rate performance in the hovering UAV scenario with different numbers of IoT devices, $N$.}
	\label{fig:NoUEs_HoveringUAV}
\end{figure}

Firstly, we consider the hovering UAV scenario and compare the performance versus the different number of IoT devices, $N$ with the number of RIS, $K=20$ in Fig. \ref{fig:NoUEs_HoveringUAV}. We take the average of over $1000$ episodes for each scheme to draw the figures. When using the H-PPO algorithm, the total expected throughput is higher than in other schemes including the ones using the H-DDPG algorithm, the RSS-HDDPG, WithoutRIS-HDDPG and WithoutRIS-HPPO technique. The results suggest that with the EH time and the RIS's reflecting coefficient optimization, the PPO algorithm is adequate irrespectively of the number of IoT devices.

\begin{figure}[t!]
	\centering
	\subfigure{\includegraphics[width=0.5\textwidth]{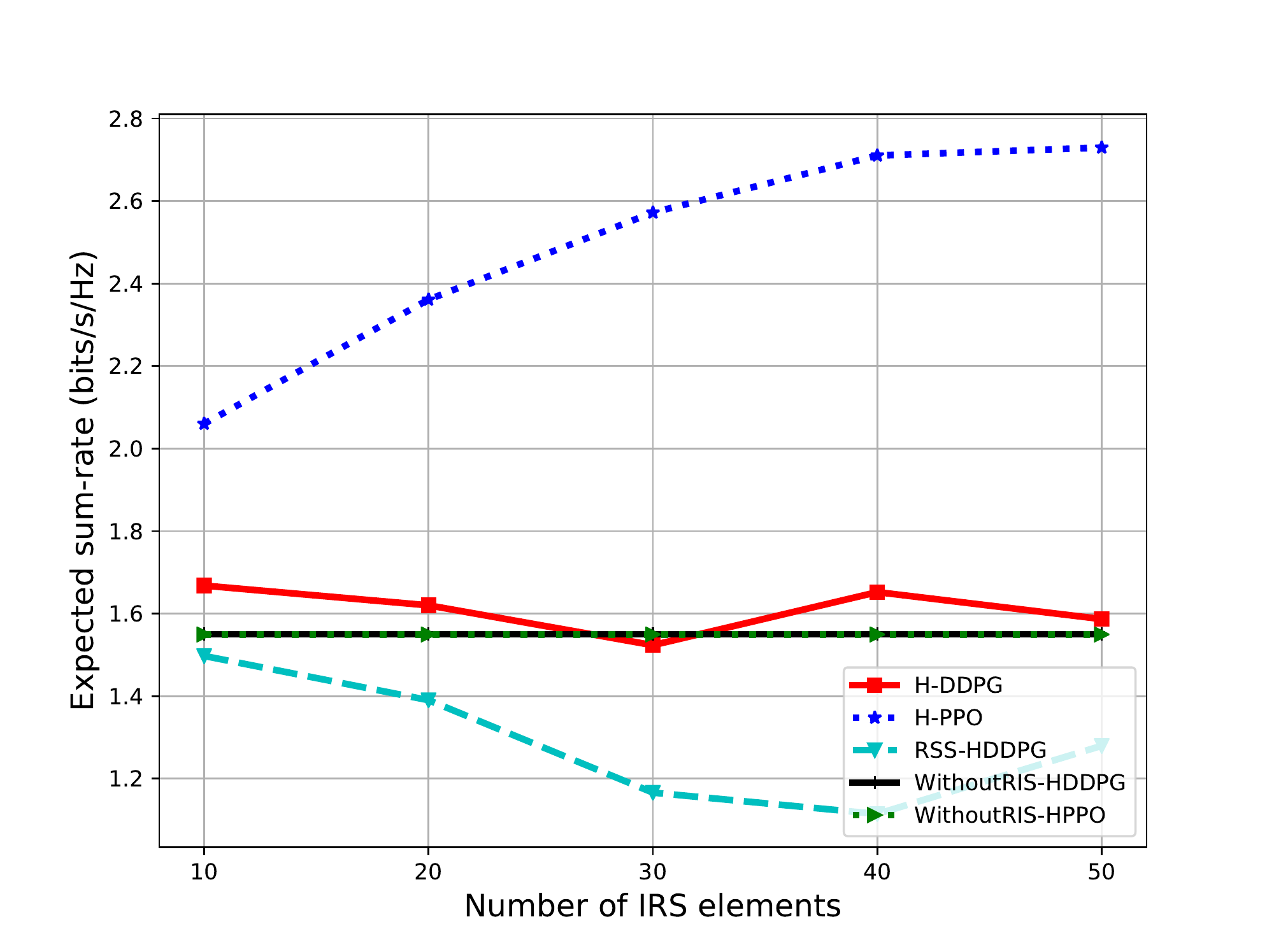}}
	\caption{The sum-rate performance  in the hovering UAV scenario with varying number of RIS elements, $K$.}
	\label{fig:NoRIS_HoveringUAV}
\end{figure}

Next, we present the achieved sum-rate of the PPO and DDPG algorithm in the hovering UAV scenario comparing with the RSS and without RIS case while the number of IoT devices is fixed at $N=10$ in Fig.~\ref{fig:NoRIS_HoveringUAV}. The H-PPO again shows the effective results with different number of RIS elements, $K$. The sum-rate performance of the H-PPO algorithm improves from $2.0$ to $2.8$ (bits/s/Hz) following the increase of the RIS elements. The sum-rate performance of the H-DDPG algorithm is slightly higher than the ones using RSS and without RIS schemes. The RIS is a passive reflector; thus, the reflected signal is diverse and not toward the destinations if we can not control the coefficient of the RIS and select the phase shift randomly. Moreover, the PPO and the DDPG algorithm reach similar results when we only optimize the EH time without the RIS.

\begin{figure}[t!]
	\centering
	\subfigure{\includegraphics[width=0.5\textwidth]{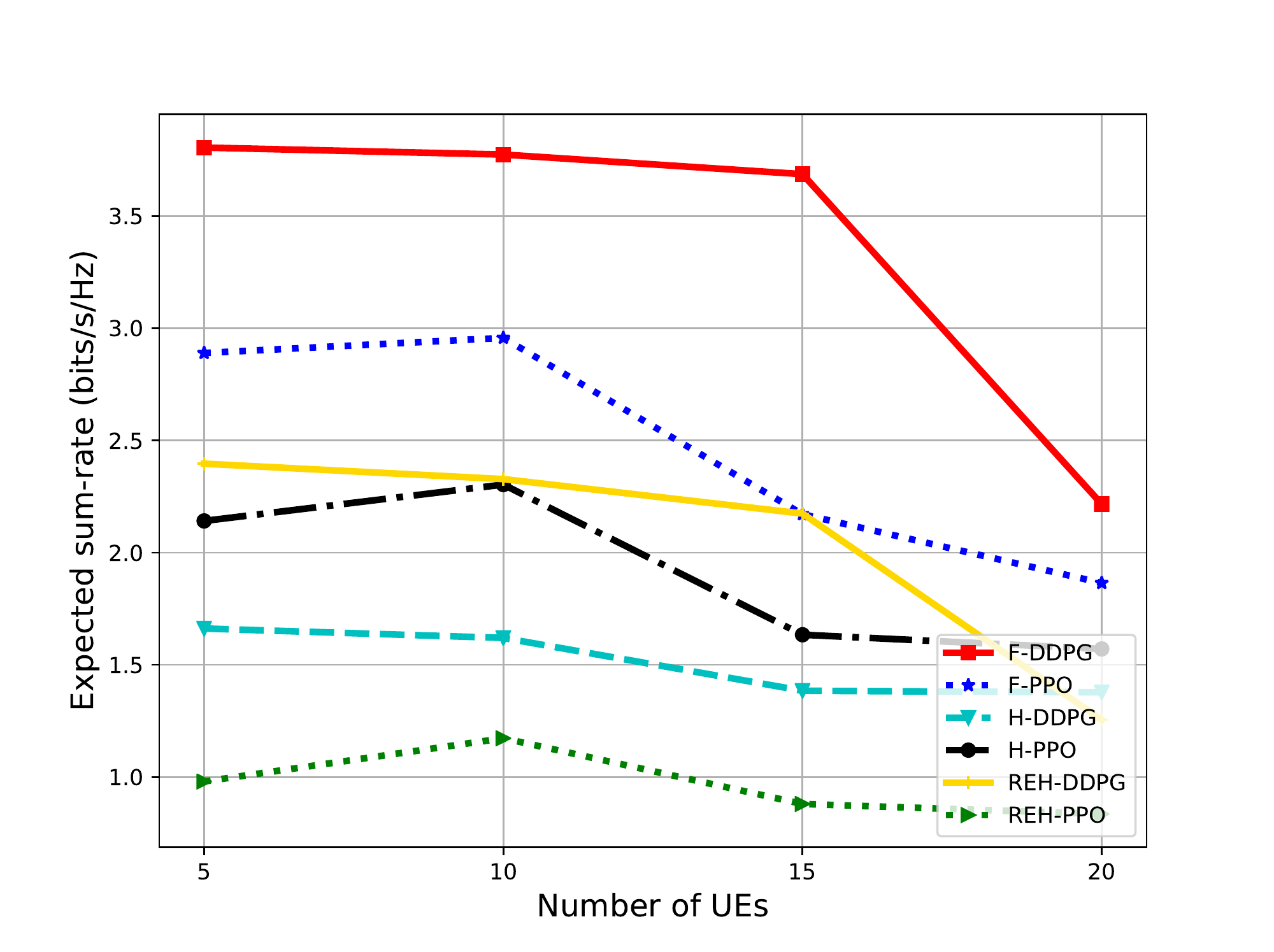}}
	\caption{The sum-rate performance with different number of IoT devices, $N$.}
	\label{fig:NoIoT}
\end{figure}

In Fig. \ref{fig:NoIoT}, we compare the total sum-rate in the mobile UAV with the number of RIS elements $K=20$ and different numbers of IoT devices, $N$. In contrast with the hovering scenarios, the method based on the F-DDPG algorithm shows impressive results over other schemes. When using the F-DDPG algorithm, we can achieve the total throughput of around $3.8$ bits/Hz. The F-PPO algorithm is not good and trapped in an optimal local value. The reason is that the F-PPO algorithm is an on-policy method and offers less random exploration over the training

\begin{figure}[t!]
	\centering
	\subfigure{\includegraphics[width=0.5\textwidth]{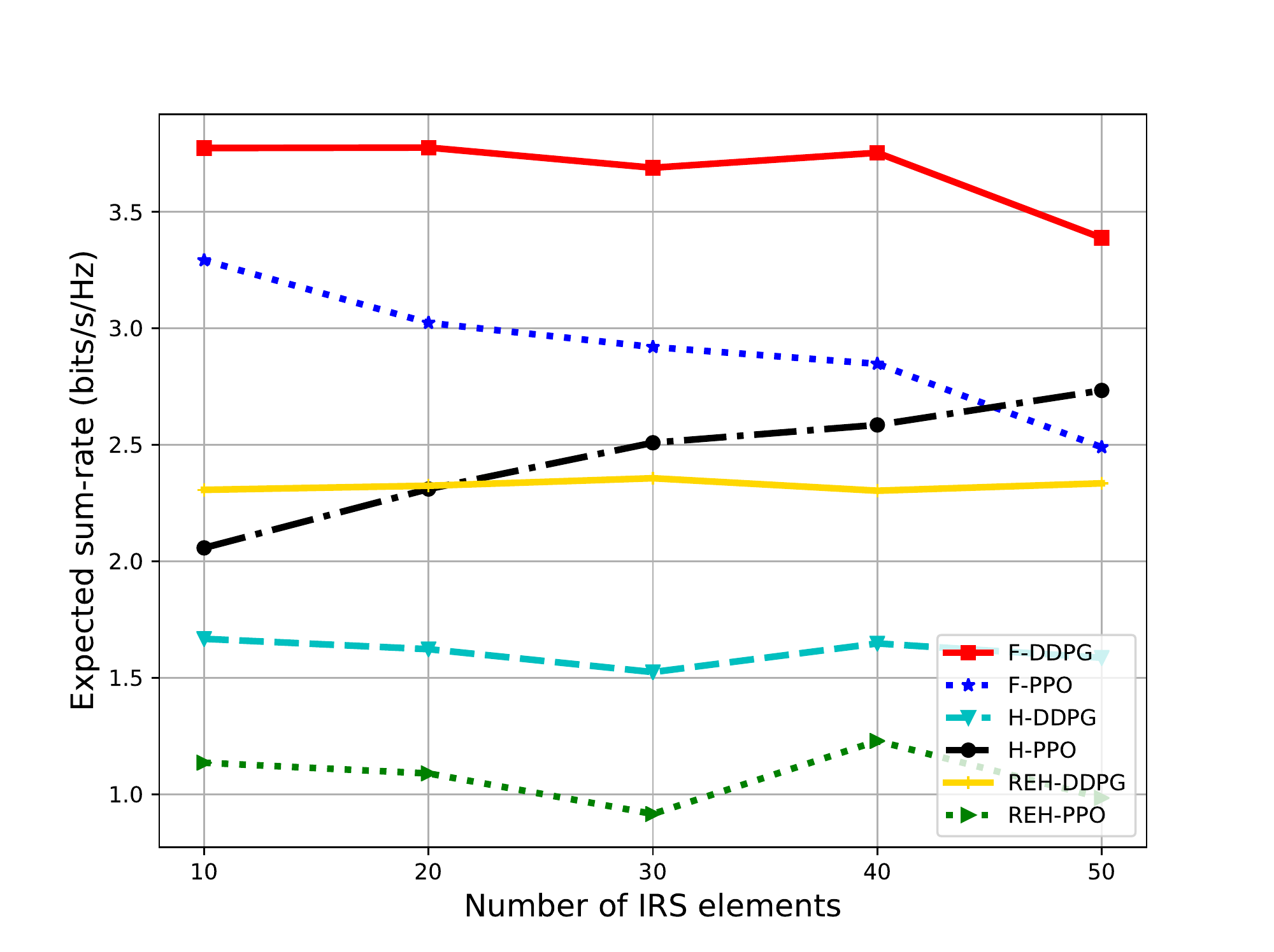}}
	\caption{The sum-rate performance with different numbers of RIS elements, $K$.}
	\label{fig:NoRIS}
\end{figure}

We consider the different number of RIS element $K$ and compare the performance of our proposed algorithms with REH schemes and hovering UAV scenario in Fig.~\ref{fig:NoRIS}. The F-DDPG algorithm-based technique outperforms other schemes while it reaches around $3.8$ (bits/s/Hz). Following the F-DDPG algorithm is the performance using the F-PPO algorithm in the mobile UAV. When we jointly optimize the UAV's trajectory, IoT's EH time and RIS's phase shift, the achievable sum-rate is significantly increased in comparison with the case when we optimize only the EH time, RIS phase shift in hovering scenario and when we consider the optimization of trajectory and EH time in REH-DDPG, REH-PPO algorithm.

\section{Conclusion}\label{Sec:Con}
In this paper, we have introduced a new system model for RIS-assisted UAV communications with the downlink power transfer and uplink information transmission protocol. By utilizing the UAV's mobility, the flexibility of the RIS, and the effectiveness of the protocol, the RIS-assisted UAV network is a promising technique for practical applications. We have proposed two DRL techniques for jointly optimizing the UAV's trajectory, IoT's EH time scheduling and the phase shift matrix of the RIS to maximize the network's throughput. The results suggest that the systems learned by the DRL algorithm can deal with dynamic environments and satisfy some power restrictions and processing time in RIS-assisted UAV communications. In the future, we plan to extend our work to include a distributed model and cooperative communications with multiple UAVs.


\bibliographystyle{IEEEtran}

\bibliography{IEEEabrv,reference}

\begin{thebibliography}{10}
\providecommand{\url}[1]{#1}
\csname url@samestyle\endcsname
\providecommand{\newblock}{\relax}
\providecommand{\bibinfo}[2]{#2}
\providecommand{\BIBentrySTDinterwordspacing}{\spaceskip=0pt\relax}
\providecommand{\BIBentryALTinterwordstretchfactor}{4}
\providecommand{\BIBentryALTinterwordspacing}{\spaceskip=\fontdimen2\font plus
\BIBentryALTinterwordstretchfactor\fontdimen3\font minus
  \fontdimen4\font\relax}
\providecommand{\BIBforeignlanguage}[2]{{%
\expandafter\ifx\csname l@#1\endcsname\relax
\typeout{** WARNING: IEEEtran.bst: No hyphenation pattern has been}%
\typeout{** loaded for the language `#1'. Using the pattern for}%
\typeout{** the default language instead.}%
\else
\language=\csname l@#1\endcsname
\fi
#2}}
\providecommand{\BIBdecl}{\relax}
\BIBdecl

\bibitem{Long:EAI}
L.~D. Nguyen, A.~Kortun, and T.~Q. Duong, ``An introduction of real-time
  embedded optimisation programming for {UAV} systems under disaster
  communication,'' \emph{{EAI} Endorsed Transactions on Industrial Networks and
  Intelligent Systems}, vol.~5, no.~17, pp. 1--8, Dec. 2018.

\bibitem{Khoi:20:Access}
K.~K. Nguyen, N.~A. Vien, L.~D. Nguyen, M.-T. Le, L.~Hanzo, and T.~Q. Duong,
  ``Real-time energy harvesting aided scheduling in {UAV}-assisted {D2D}
  networks relying on deep reinforcement learning,'' \emph{IEEE Access},
  vol.~9, pp. 3638--3648, Dec. 2021.

\bibitem{KK:21:TCOM}
\BIBentryALTinterwordspacing
K.~K. Nguyen, T.~Q. Duong, T.~Do-Duy, H.~Claussen, and L.~Hanzo, ``{3D UAV}
  trajectory and data collection optimisation via deep reinforcement
  learning,'' 2021. [Online]. Available: \url{https://arxiv.org/abs/2106.03129}
\BIBentrySTDinterwordspacing

\bibitem{KK:19:Access}
K.~K. Nguyen, T.~Q. Duong, N.~A. Vien, N.-A. Le-Khac, and L.~D. Nguyen,
  ``Distributed deep deterministic policy gradient for power allocation control
  in {D2D}-based {V2V} communications,'' \emph{IEEE Access}, vol.~7, pp.
  164\,533--164\,543, Nov. 2019.

\bibitem{Khoi:19:Access}
K.~K. Nguyen, T.~Q. Duong, N.~A. Vien, N.-A. Le-Khac, and N.~M. Nguyen,
  ``Non-cooperative energy efficient power allocation game in {D2D}
  communication: A multi-agent deep reinforcement learning approach,''
  \emph{IEEE Access}, vol.~7, pp. 100\,480--100\,490, Jul. 2019.

\bibitem{LN:19:SPAWC}
L.~D. Nguyen, K.~K. Nguyen, A.~Kortun, and T.~Q. Duong, ``Real-time deployment
  and resource allocation for distributed {UAV} systems in disaster relief,''
  in \emph{Proc. IEEE 20th International Workshop on Signal Processing Advances
  in Wireless Commun. (SPAWC)}, Cannes, France, Jul. 2019, pp. 1--5.

\bibitem{EB:19:Access}
E.~Basar, M.~D. Renzo, J.~D. Rosny, M.~Debbah, M.-S. Alouini, and R.~Zhang,
  ``Wireless communications through reconfigurable intelligent surfaces,''
  \emph{IEEE Access}, vol.~7, pp. 116\,753--116\,773, Aug. 2019.

\bibitem{SA:20:TC}
S.~Atapattu, R.~Fan, P.~Dharmawansa, G.~Wang, J.~Evans, and T.~A. Tsiftsis,
  ``Reconfigurable intelligent surface assisted two–way communications:
  Performance analysis and optimization,'' \emph{IEEE Trans. Commun.}, vol.~68,
  no.~10, pp. 6552--6567, Oct. 2020.

\bibitem{HY:20:JSAC}
H.~Yu, H.~D. Tuan, A.~A. Nasir, T.~Q. Duong, and H.~V. Poor, ``Joint design of
  reconfigurable intelligent surfaces and transmit beamforming under proper and
  improper {Gaussian} signaling,'' \emph{IEEE J. Select. Areas Commun.},
  vol.~38, no.~11, pp. 2589--2603, Nov. 2020.

\bibitem{YL:21:TC}
Y.~Li, M.~Jiang, Q.~Zhang, and J.~Qin, ``Joint beamforming design in
  multi-cluster {MISO NOMA} reconfigurable intelligent surface-aided downlink
  communication networks,'' \emph{IEEE Trans. Commun.}, vol.~69, no.~1, pp.
  664--674, Jan. 2021.

\bibitem{LG:20:Access}
L.~Ge, P.~Dong, H.~Zhang, J.-B. Wang, and X.~You, ``Joint beamforming and
  trajectory optimization for intelligent reflecting surfaces-assisted {UAV}
  communications,'' \emph{IEEE Access}, vol.~8, pp. 78\,702--78\,712, Apr.
  2020.

\bibitem{SL:20:WCL}
S.~Li, B.~Duo, X.~Yuan, Y.-C. Liang, and M.~D. Renzo, ``Reconfigurable
  intelligent surface assisted {UAV} communication: Joint trajectory design and
  passive beamforming,'' \emph{IEEE Wireless Commun. Lett.}, vol.~9, no.~5, pp.
  716--720, May 2020.

\bibitem{AR:21:IOT}
A.~Ranjha and G.~Kaddoum, ``{URLLC} facilitated by mobile {UAV} relay and
  {RIS}: A joint design of passive beamforming, blocklength, and {UAV}
  positioning,'' \emph{{IEEE} Internet Things J.}, vol.~8, no.~6, pp.
  4618--4627, Mar. 2021.

\bibitem{KK:21:NCE}
\BIBentryALTinterwordspacing
K.~K. Nguyen, S.~Khosravirad, L.~D. Nguyen, T.~T. Nguyen, and T.~Q. Duong,
  ``Intelligent reconfigurable surface-assisted multi-{UAV} networks: Efficient
  resource allocation with deep reinforcement learning,'' 2021. [Online].
  Available: \url{https://arxiv.org/abs/2105.14142}
\BIBentrySTDinterwordspacing

\bibitem{YZ:20:VT}
Y.~Zou, S.~Gong, J.~Xu, W.~Cheng, D.~T. Hoang, and D.~Niyato, ``Wireless
  powered intelligent reflecting surfaces for enhancing wireless
  communications,'' \emph{{IEEE} Trans. Veh. Technol.}, vol.~69, no.~10, pp.
  12\,369--12\,373, Oct. 2020.

\bibitem{CH:20:JSAC}
C.~Huang, R.~Mo, and C.~Yuen, ``Reconfigurable intelligent surface assisted
  multiuser {MISO} systems exploiting deep reinforcement learning,'' \emph{IEEE
  J. Select. Areas Commun.}, vol.~38, no.~8, pp. 1839--1850, Aug. 2020.

\bibitem{Pan:20:JSAC}
C.~Pan \emph{et~al.}, ``Intelligent reflecting surface aided {MIMO}
  broadcasting for simultaneous wireless information and power transfer,''
  \emph{IEEE J. Select. Areas Commun.}, vol.~38, no.~8, pp. 1719--1734, Aug.
  2020.

\bibitem{HY:21:SP}
H.~Yang, X.~Yuan, J.~Fang, and Y.-C. Liang, ``Reconfigurable intelligent
  surface aided constant-envelope wireless power transfer,'' \emph{IEEE Trans.
  Signal Process.}, vol.~69, pp. 1347--1361, Feb. 2021.

\bibitem{SL:21:WC}
S.~Lin, B.~Zheng, G.~C. Alexandropoulos, M.~Wen, M.~D. Renzo, and F.~Chen,
  ``Reconfigurable intelligent surfaces with reflection pattern modulation:
  Beamforming design and performance analysis,'' \emph{IEEE Trans. Wireless
  Commun.}, vol.~20, no.~2, pp. 741--754, Feb. 2021.

\bibitem{LW:20:arvix}
\BIBentryALTinterwordspacing
L.~Wang, K.~Wang, C.~Pan, W.~Xu, and N.~Aslam, ``Joint trajectory and passive
  beamforming design for intelligent reflecting surface-aided {UAV}
  communications: A deep reinforcement learning approach,'' 2020. [Online].
  Available: \url{https://arxiv.org/abs/2007.08380}
\BIBentrySTDinterwordspacing

\bibitem{HY:21:WC}
H.~Yang, Z.~Xiong, J.~Zhao, D.~Niyato, L.~Xiao, and Q.~Wu, ``Deep reinforcement
  learning-based intelligent reflecting surface for secure wireless
  communications,'' \emph{IEEE Trans. Wireless Commun.}, vol.~20, no.~1, pp.
  375--388, Jan. 2021.

\bibitem{KF:20:WCL}
K.~Feng, Q.~Wang, X.~Li, and C.-K. Wen, ``Deep reinforcement learning based
  intelligent reflecting surface optimization for {MISO} communication
  systems,'' \emph{IEEE Wireless Commun. Lett.}, vol.~9, no.~5, pp. 745--749,
  May 2020.

\bibitem{YC:20:WC}
Y.~Chen, B.~Ai, H.~Zhang, Y.~Niu, L.~Song, Z.~Han, and H.~V. Poor,
  ``Reconfigurable intelligent surface assisted device-to-device
  communications,'' \emph{IEEE Trans. Wireless Commun.}, vol.~20, no.~5, pp.
  2792--2804, May 2021.

\bibitem{BD:95:Book:v1}
D.~P. Bertsekas, \emph{Dynamic Programming and Optimal Control}.\hskip 1em plus
  0.5em minus 0.4em\relax Athena Scientific Belmont, MA, 1995, vol.~1, no.~2.

\bibitem{Lillicrap:15}
T.~P. Lillicrap \emph{et~al.}, ``Continuous control with deep reinforcement
  learning,'' in \emph{Proc. 4th International Conf. on Learning
  Representations (ICLR)}, 2016.

\bibitem{JS:16:ICLR}
J.~Schulman, P.~Moritz, S.~Levine, M.~I. Jordan, and P.~Abbeel,
  ``High-dimensional continuous control using generalized advantage
  estimation,'' in \emph{Proc. 4th International Conf. Learning Representations
  (ICLR)}, 2016.

\bibitem{JS:17:PPO}
\BIBentryALTinterwordspacing
J.~Schulman, F.~Wolski, P.~Dhariwal, A.~Radford, and O.~Klimov, ``Proximal
  policy optimization algorithms,'' 2017. [Online]. Available:
  \url{https://arxiv.org/abs/1707.06347}
\BIBentrySTDinterwordspacing

\bibitem{Mnih:16}
V.~Mnih \emph{et~al.}, ``Asynchronous methods for deep reinforcement
  learning,'' in \emph{Proc. Int. Conf. Mach. Learn.}\hskip 1em plus 0.5em
  minus 0.4em\relax PMLR, 2016, pp. 1928--1937.

\bibitem{Abadi:16}
M.~Abadi \emph{et~al.}, ``{Tensorflow}: A system for large-scale machine
  learning,'' in \emph{Proc. 12th USENIX Sym. Opr. Syst. Design and Imp. (OSDI
  16)}, Nov. 2016, pp. 265--283.

\end{thebibliography}

\end{document}